\documentclass[twocolumn,showpacs]{revtex4}
\usepackage{graphicx}
\usepackage{dcolumn}

\begin{document}


\title{Unique features of structure in an odd-proton $N \approx Z$ nucleus $^{69}$As}

\author{M. Hasegawa,$^{1}$ K. Kaneko,$^{2}$ and
        T. Mizusaki,$^{3}$}

\affiliation{
$^{1}$Laboratory of Physics, Fukuoka Dental College, Fukuoka 814-0193, Japan \\
$^{2}$Department of Physics, Kyushu Sangyo University, Fukuoka 813-8503, Japan \\
$^{3}$Institute of Natural Sciences, Senshu University, Tokyo 101-8425, Japan
}

\date{\today}

\begin{abstract}

   We apply a large-scale shell model to the study of proton-rich
odd-mass nuclei with $N \approx Z$.
Calculations predict unexpected structure in the $^{69}$As nucleus
for which a detailed experiment was recently performed.
In this odd-proton nucleus, one neutron competes with one proton
for occupying the high-$j$ intruder orbit $g_{9/2}$ in the $9/2_1^+$ state
and almost solely occupies the $g_{9/2}$ orbit in other low-lying
positive-parity states.
The $T=0$, $J=9$ one-proton-one-neutron alignment takes place
in the negative-parity states with medium-high spins.
A unique coexistence of two lowest bands with positive and negative signs
of spectroscopic quadrupole moments is predicted.

\end{abstract}

\pacs{21.60.Cs,21.10.-k,27.50.+e}

\maketitle

\section{Introduction}\label{sec1}

   The study of $N \approx Z$ proton-rich nuclei with $A=60-100$ is
 one of current topics in nuclear physics. These nuclei have provided
 plenty of phenomena exhibiting interesting structure
 such as rapid shape changes with increasing $N$, $Z$, and $J$ (spin). 
 Strong residual correlations between protons and neutrons
 which are inherent properties of nuclei play a key role
 in these $N \approx Z$ $fpg$-shell nuclei.
 Experiments are extensively accumulating a variety of unique phenomena
 which demand theoretical explanations.
  The shape changes, shape coexistence, and particle alignments
 have been investigated intensively in even-even $N \approx Z$ nuclei
 until now (for instance , see latest work 
 \cite{Fischer1,Bouchez,Fischer2,Stefanova,Sarrig,Petrovici,Kobayasi,Sun,Almehed}), 
 but they have not fully been discussed in odd-mass $N \approx Z$ nuclei and
 odd-odd $N \approx Z$ nuclei  \cite{Bruce,Jenkins}.
  One of the reasons apart from experimental difficulty
 is that theoretical approaches employed are not necessarily fit
 to describe these properties of odd-mass and odd-odd nuclei.
 
   Recently, the spherical shell model has become capable of calculating
 $A=60-70$ nuclei in the $pf_{5/2}g_{9/2}$ shell. The shell model calculations
 reproduce observed energy levels up to high spins in these nuclei
 and successfully explain shape changes from $^{64}$Ge to $^{68}$Se
 along the even-even $N=Z$ line and various particle alignments
 in even-even nuclei \cite{Hase1,Kaneko1,Hase2}.
 The shell model predicts unexpected alignment in this mass region.
 More precisely, $T=0$ one-proton-one-neutron ($1p1n$) alignment
 in the high-$j$ $g_{9/2}$ orbit
 takes place not only in odd-odd nuclei but also in even-even nuclei.
 In addition, the shell model calculation \cite{Hase3} revealed possible
 shape difference between low-lying states in the odd-odd $N=Z$ nucleus $^{66}$As,
 predicting a shape isomer. These calculations suggest that unusual structure
 observed in even-even $N \approx Z$ nuclei would occur in odd-mass and
 odd-odd nuclei of this mass region and the reverse also could happen.
 A further study of the questions must contribute to elucidation
 of various features of nuclear structure.
 
    In this paper, we are interested in the question what characteristics
 appear in odd-mass $N \approx Z$ nuclei. Our previous paper \cite{Hase3},
 where the isomeric state $9/2_1^+$ in $^{67}$As was investigated,
 found unique structure formed in the odd-proton $N \approx Z$ nucleus.
 A recent experiment carried out by Stefanescu {\it et al.} \cite{Stefanescu}
 has provided much new information in the nucleus $^{69}$As as compared with
 nearby other odd-mass nuclei. Three doublets of negative- and positive-parity
 bands and high-spin states up to band terminations are observed in $^{69}$As.
 The very detailed data seem to conceal unknown features of odd-proton
 $pfg$-shell nuclei and are worth investigating.
 The shell model calculation which gives precise wave functions
 for yrast and non-yrast states is expected to be fruitful for the investigation.

\section{The model}\label{sec2}

   We employ the same shell model with the single-particle states
$(p_{3/2},f_{5/2},p_{1/2},g_{9/2})$ as that used for $^{66}$As and $^{67}$As
in Ref. \cite{Hase3}.  The extended $P+QQ$ Hamiltonian used here is composed of
the single-particle energies, $T=0$ monopole field ($H^{T=0}_{\pi \nu}$),
monopole corrections ($H_{\rm mc}$),
$J=0$ pairing force ($H_{P_0}$), quadrupole-quadrupole force ($H_{QQ}$)
and octupole-octupole force ($H_{OO}$):
\begin{eqnarray}
 H & = & H_{\rm sp} + H^{T=0}_{\pi \nu} + H_{\rm mc}
        + H_{P_0} + H_{QQ} + H_{OO}  \nonumber \\
   & = & \sum_{\alpha} \varepsilon_a c_\alpha^\dag c_\alpha
       - k^0 \sum_{a \leq b} \sum_{JM} A^\dagger_{JM00}(ab) A_{JM00}(ab)
            \nonumber \\
   & + & \sum_{a \leq b} \sum_{T} \Delta k_{\rm mc}^T(ab) \sum_{JMK}
               A^\dagger_{JMTK}(ab) A_{JMTK}(ab)         \nonumber \\
   & - & \frac{1}{2} g_0 \sum_K P^\dag_{001K} P_{001K}\nonumber \\
   & - & \frac{1}{2} \frac{\chi_2}{b^4}  \sum_M :Q^\dag_{2M} Q_{2M}:
         - \frac{1}{2} \frac{\chi_3}{b^6} \sum_M :O^\dag_{3M} O_{3M}:,
          \label{eq:1}
\end{eqnarray}
where $A^\dagger_{JMTK}(ab)$ is a pair creation  operator
with spin $JM$ and isospin $TK$ in the orbits ($ab$).
This shell model has proven to be rather successful in describing $A=64-70$ nuclei
such as $^{65}$Ge, $^{67}$Ge, $^{66}$As, $^{67}$As, 
and neighboring even-even nuclei up to $^{70}$Ge.
For more details, see Refs.  \cite{Hase1,Kaneko1,Hase2,Hase3}.
Some specifics are reviewed below.

   We made a search for good parameters which describe even- and odd-mass
Ge isotopes \cite{Hase2}.  The single-particle energies of the four orbits
are obtained as
 $\varepsilon_{p3/2} = 0.0$, $\varepsilon_{f5/2} = 0.77$,
 $\varepsilon_{p1/2} = 1.11$, and $\varepsilon_{g9/2} = 2.50$
 in MeV.
The value $\varepsilon_{g9/2}=2.5$ MeV differs from 3.7 MeV extracted from
$^{57}$Ni in Ref. \cite{Rudolph} but coincides with that of Ref. \cite{Aberg}.
It is difficult to reproduce the energy levels ($9/2^+$ and others)
of the odd-mass Ge isotopes $^{65}$Ge and $^{67}$Ge
unless we lower the energy $\varepsilon_{g9/2}$ to 2.5 MeV.
 The monopole field $H^{T=0}_{\pi \nu}$, which depends only on the total isospin
and the number of valence nucleons, does not affect excitation energies
considered in this paper.
Because the isospin is a good quantum number in $N \approx Z$ nuclei,
we set the Hamiltonian to conserve the isospin.
  The parameter search gave the force strengths
($g_0 = 0.27(64/A)$, $\chi_2 = 0.25(64/A)^{5/3}$, and $\chi_3 = 0.05(64/A)^2$
 in MeV) and five parameters of the monopole corrections.
 Three of the monopole corrections
 ($\Delta k^{T=1}_{\rm mc}(p_{3/2},f_{5/2}) =
  \Delta k^{T=1}_{\rm mc}(p_{3/2},p_{1/2}) = -0.3$ and
  $\Delta k^{T=1}_{\rm mc}(f_{5/2},p_{1/2}) = -0.4$ in MeV)
are important for the collectivity in the $(p_{3/2},f_{5/2},p_{1/2})$ subspace.
Especially, $\Delta k^{T=1}_{\rm mc}(f_{5/2},p_{1/2})$ plays an important role
in producing the oblate shape for $^{68}$Se and $^{72}$Kr \cite{Kaneko1}.
   The study of isomeric states in $^{66}$As and $^{67}$As
compelled us to make a modification to the monopole corrections \cite{Hase3}.
The additional terms are
$\Delta k^{T=0}_{\rm mc}(a,g_{9/2}) = -0.18$ MeV ($a=p_{3/2},f_{5/2},p_{1/2}$).
These corrections have an effect of lowering $\varepsilon_{g9/2}$ further
but the effect is different from wide effects caused by directly decreasing 
$\varepsilon_{g9/2}$. 
 Our calculations suggest that a special contribution of the $g_{9/2}$ orbit
may be related to the rapid structure change when $Z$ and $N$ approaching 40, 
although the parameter-fitting treatment does not explain the physical origin.

   In this paper, we carry out large-scale shell model calculations
for $^{69}$As using the code \cite{Mizusaki1}, which diagonalizes 
shell model Hamiltonian matrix in $M$-scheme.
As, in $M$-scheme, shell model dimensions decrease as a function 
of $|M|$ value, lower spin state needs bigger shell model dimension
and there is a case that higher spin states are easily solved
while the lower spin states can not be solved exactly.
In the present case, the low-spin states with
 $J^\pi \le 7/2^-$ or $J^\pi \le 9/2^+$ in $^{69}$As
have a dimension of the Hamiltonian matrix larger than $2 \times 10^8$.
The dimension exceeds the capacity of our personal computers.
Therefore, we truncate the matrix by limiting the number $t$ in 
the configuration space
 $\Sigma_{n\ge t}(p_{3/2},f_{5/2})^n(p_{1/2},g_{9/2})^{13-n}$.
For $3/2^-$ states, for instance, the dimension is $1.996 \times 10^8$
when $t=5$. For $^{69}$As, however, the maximum dimension is at most
$2.4 \times 10^8$ and hence the $t=5$ truncation ($t=2$ for $9/2^+$)
is expected to be good enough for our discussions.

For more accurate calculations, we consider the energy variance extrapolation
method \cite{Mizusaki2}, which can evaluate exact energy from the truncated 
wave functions based on the scaling property between energy difference
$\delta E$ and energy variance $\Delta E$, {\i.e.,} $\delta E \propto \Delta E$.
Here we use a new formula of energy variance extrapolation derived by introducing 
a lowest energy projection operator $\hat R$ to approximated wave function 
$\left| \varphi  \right\rangle $.
As $\hat R\left| \varphi  \right\rangle$ is a better wave function, 
it gives more precise estimation.
In fact, we use $\sqrt H$ as $\hat R$ and a new scaling relation can be 
derived, which involves the expectation value of $H^3$ \cite{Mizusaki-finustar}.
In the case of $^{69}$As, this extrapolation gives exact energies
within the accuracy 1 keV. We have found that the results in the slightly
truncated spaces for the states with $J^\pi \le 7/2^-$ or $J^\pi \le 9/2^+$
are almost exact. The extrapolation improves energies only by several keV.

We adopt the harmonic-oscillator range parameter $b \approx 1 \times A^{1/6}$ fm.
We use the effective charges $e_{\rm eff}^\pi=1.5e$ and $e_{\rm eff}^\nu=0.5e$ 
for $B(E2)$ and the spectroscopic quadrupole moment
$Q_{\rm sp}= \Sigma_{\tau = \pi \nu} e_{\rm eff}^\tau \sqrt{16\pi/5}
 \langle r^2 Y_{20} \rangle_\tau$,
where the symbol $\langle \ \ \rangle$ denotes an expectation value.

\section{Calculated results and discussions}\label{sec3}

\subsection{Positive-parity states}\label{sec3.1}

\begin{figure}[b]
\includegraphics[width=8.6cm,height=12.0cm]{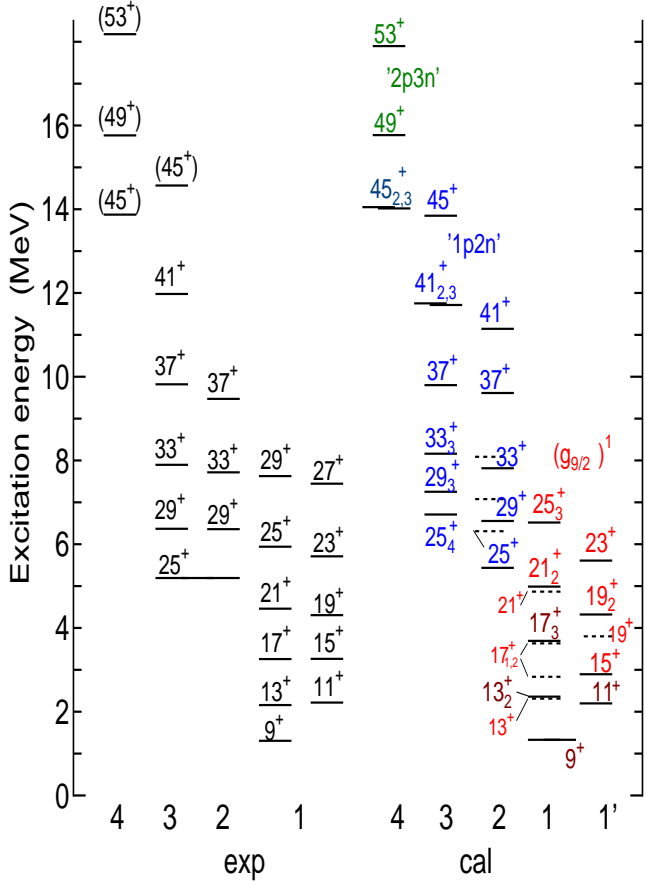}
  \caption{Experimental and calculated energy levels with positive parity.
           The spin of each state is denoted by the double number $2J$.
           Non-yrast states are distinguished with the subscripts
           which represent the numerical order for each spin,
           while the subscript 1 is omitted for the yrast states.
           In addition to these ``collective states",
           nearby levels
           which have the same spin as that of the collective states
           are plotted using the dashed lines. }
  \label{fig1}
\end{figure}

   Calculated energy levels with positive parity are compared with
the experimental ones of Ref. \cite{Stefanescu}, in Fig. \ref{fig1}.
The experimental bands are labeled with the same numbers
as those used in Ref. \cite{Stefanescu}, where these bands are classified
according to observed electromagnetic transitions.
Similarly, we classify the calculated energy levels into bands
according to cascade series of large $B(E2:J \rightarrow J-2)$ values.
 It should be noted that the calculated $B(E2:J \rightarrow J-2)$ value
is very small when the two states with $J-2$ and $J$ belong to different bands.
  We calculated also $B(M1:J \rightarrow J-1)$ values to compare with
experimentally observed $J \rightarrow J-1$ transitions.
We use the free-nucleon effective $g$-factors for calculations of $M1$
and $M2$ transition strengths.
In the calculations, the pairs of ($13/2_1^+$, $13/2_2^+$),
 ($17/2_2^+$, $17/2_3^+$), and ($21/2_1^+$, $21/2_2^+$)
almost degenerate in energy, respectively.
We classify the $21/2_2^+$ state as the member of the band 1
from the value $B(E2:25/2_3^+ \rightarrow 21/2_2^+)=12.7$ Weisskoph unit (W.u.)
much larger than $B(E2:25/2_3^+ \rightarrow 21/2_1^+)=0.15$ W.u.
This is consistent with the calculated values 
$B(M1:23/2_1^+ \rightarrow 21/2_2^+) \gtrsim B(M1:23/2_1^+ \rightarrow 21/2_1^+)
\sim 0.02$ W.u., $B(E2:23/2_1^+ \rightarrow 21/2_2^+)=1.7$ W.u.,
and $B(E2:23/2_1^+ \rightarrow 21/2_1^+)=0.03$ W.u.
For $13/2^+$, we have the values 
$B(M1:15/2_1^+ \rightarrow 13/2_1^+)=2.96$ W.u.,
$B(M1:15/2_1^+ \rightarrow 13/2_2^+)=0.95$ W.u.,
$B(E2:15/2_1^+ \rightarrow 13/2_1^+)=12.6$ W.u.,
and $B(E2:15/2_1^+ \rightarrow 13/2_2^+)=0.70$ W.u.
These values suggest that the $13/2_1^+$ state is another candidate
for the $13/2^+$ state observed in the band 1,
although the $B(E2)$ values for the cascade decay
 $17/2_3^+ \rightarrow 13/2_1^+ \rightarrow 9/2_1^+$ are smaller than
 those for $17/2_3^+ \rightarrow 13/2_2^+ \rightarrow 9/2_1^+$.
We get calculated values $B(E2:11/2_1^+ \rightarrow 9/2_1^+)=7.3$ W.u. 
and $B(M1:11/2_1^+ \rightarrow 9/2_1^+)=0.22$ W.u.
This indicates a strong $E2$ transition for $11/2_1^+ \rightarrow 9/2_1^+$.

   We cannot find the $27/2^+$ ($29/2^+$) member belonging to
the band 1 ($1^\prime$) among the lowest five states with $J=27/2$ ($J=29/2$)
in calculations.  These states with $J^\pi \ge 27/2^+$ have
different configurations from the states with $J^\pi \le 23/2^+$ 
and the $25/2_3^+$ state as discussed later.
Therefore, the $27/2^+$ and $29/2^+$ members are missing
in the calculated band 1 ($1^\prime$) in Fig. \ref{fig1}. 
Our model predicts the $25/2_4^+$ state as a member of the band 3.
The calculated results thus show deviations from the experimental
data \cite{Stefanescu} and suggest that mixing of wave functions
should be improved.
However, our model reproduces qualitatively well
a lot of observed energy levels
and basically explains the experimental band classifications.
It is difficult to get such a good quality of agreement with experiment
using other models at present.
The structure study of $^{69}$As with our model is worthwhile.

   Let us analyze obtained wave functions
and consider the structure which characterizes the observed bands.
We calculated the following quantities for every state:
(1) expectation values of proton and neutron numbers in the four orbits
which are denoted by $\langle n^\pi_a \rangle $ and $\langle n^\nu_a \rangle $;
(2) expectation values of spin and isospin in the high-$j$ intruder orbit
$g_{9/2}$ and in the $pf$ shell ($p_{3/2}$,$f_{5/2}$,$p_{1/2}$)
which are evaluated as
 $\langle J_i \rangle=[\langle {\hat j_i}^2 \rangle +1/4]^{1/2}-1/2$
and $\langle T_i \rangle=[\langle {\hat t_i}^2 \rangle +1/4]^{1/2}-1/2$,
 where ${\hat j_i}$ and ${\hat t_i}$ are spin and isospin operators
 for $i=g_{9/2}$ or $i=pf$;
(3) spectroscopic quadrupole moment $Q_{\rm sp}$.
 These quantities are useful in investigating structure change
 in a series of states
\cite{Hase1,Kaneko1,Hase2,Hase3}. 

\begin{figure}[b]
\includegraphics[width=8.2cm,height=11.5cm]{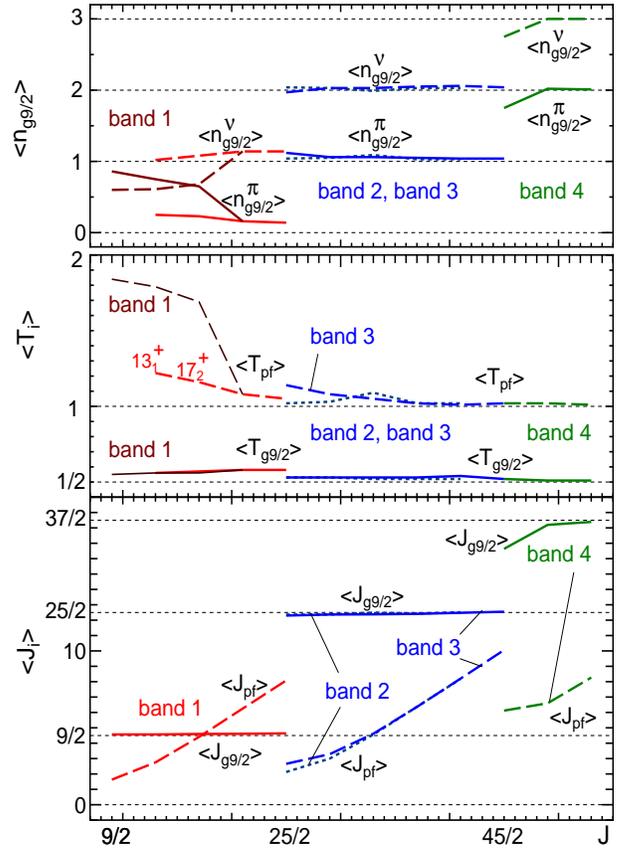}
  \caption{The expectation values $\langle n^\nu _{g9/2} \rangle$
           and $\langle n^\pi _{g9/2} \rangle$ in the upper panel,
           and the expectation values $\langle T_i \rangle$ and
           $\langle J_i \rangle$ ($i=g_{9/2}$, $pf$) in the lower panel,
           for the positive-parity bands.}
  \label{fig2}
\end{figure}

   In Fig. 2, we illustrate the expectation values $\langle n_{g9/2}^\pi \rangle$,
$\langle n_{g9/2}^\nu \rangle$, $\langle J_i \rangle$,
 and $\langle T_i \rangle$ ($i=g_{9/2}$, $pf$) for the positive-parity bands.
 Figure 2 indicates that the classifications of bands
correspond to different configurations.
As the spin $J$ increases, the positive-parity states pass 
three stages A, B, and C classified by the occupation number
 $\langle n_{g9/2}^\pi \rangle + \langle n_{g9/2}^\nu \rangle$:
[Stage A] states with $J^\pi \le 23/2^+$ and the $25/2_3^+$ state
have $\langle n_{g9/2}^\pi \rangle + \langle n_{g9/2}^\nu \rangle < 1.5$;
[Stage B] states with $25/2^+ \le J^\pi \le 41/2^+$ and the $45/2_1^+$ state
have  $\langle n_{g9/2}^\pi \rangle \approx 1$
 and $\langle n_{g9/2}^\nu \rangle \approx 2$;
[Stage C] states with $49/2^+ \le J^\pi \le 53/2^+$ and the $45/2_{2,3}^+$ states
have $\langle n_{g9/2}^\pi \rangle \approx 2$ and
 $\langle n_{g9/2}^\nu \rangle \approx 3$.
(Note that a basis state with positive parity has an odd number of
 $g_{9/2}$ nucleons.)
The band 1 is at the stage A
(which is composed of two series of the favored states with $J=9/2+2l$
 ($\alpha =9/2$) and unfavored states with $J=7/2+2l$ ($\alpha =9/2-1$),
where $l$ being an integer), 
the bands 2 and 3 are at the stage B, and the band 4 is at the stage C.
These results for the bands 2, 3, and 4 are consistent with
the results of cranked Nilsson-Strutinsky calculations in Ref. \cite{Stefanescu}.
 The lower panel of Fig. \ref{fig2} displays angular momentum alignments.
In the bands 2 and 3, one proton and two neutrons ($1p2n$) in the $g_{9/2}$
orbit align their spins to be $\langle J_{g9/2}\rangle \approx 25/2$
($9/2+9/2+7/2$) and fold their isospins to be minimum
$\langle T_{g9/2}\rangle \approx 1/2$.
In the band 4, two protons and three neutrons ($2p3n$)
in the $g_{9/2}$ orbit align their spins
 to be $\langle J_{g9/2}\rangle \approx 37/2$
 and fold their isospins to be
$\langle T_{g9/2}\rangle \approx 1/2$.
Thus, the experimental bands 2 and 3 are the $1p2n$ aligned bands
and the band 4 is the $2p3n$ aligned band.

   The structure of the lowest positive-parity band 1 was not much discussed
in Ref. \cite{Stefanescu}.
Our calculations reveal unique structure in these states.
According to the ENSDF data \cite{ENSDF}, the lowest positive-parity state
$9/2_1^+$ has a long life-time $t_{1/2}=1.35(4)$ ns, which reminds us about
the isomeric state $9/2_1^+$ of $^{67}$As ($t_{1/2}=12(2)$ ns).
For the $9/2_1^+$ state of $^{69}$As, we obtain the expectation values
 $\langle n_{g9/2}^\pi \rangle =0.86$,
$\langle n_{g9/2}^\nu \rangle = 0.60$,
 $\langle J_{g9/2}\rangle \approx 9/2$, and
 $\langle T_{g9/2}\rangle \approx 1/2$ which are similar to those of
the $9/2_1^+$ state in $^{67}$As (see Ref. \cite{Hase3}).
This result suggests that the $9/2_1^+$ state is not simply expressed as
$(g_{9/2}^\pi) \otimes ^{68}$Ge but has a significant component
of $(g_{9/2}^\nu) \otimes ^{68}$As,
indicating that one neutron competes with one proton
for jumping in the high-$j$ orbit $g_{9/2}$ in this odd-proton nucleus.
The unique structure of the $9/2_1^+$ state retards $E3$ and $M2$ transitions
to the lower negative-parity states,
{\it i.e.,} $B(E3:9/2_1^+ \rightarrow 3/2^-,5/2^-,7/2^-) < 0.08$
and $B(M2:9/2_1^+ \rightarrow 5/2^-,7/2^-) < 0.38$ in W.u.,
because the negative-parity states with $J^\pi \le 7/2^-$ below the $9/2_1^+$
state are collective states strongly mixed in the $pf$ space as discussed later.
The $5/2^+$ and $7/2^+$ states are above the $9/2_1^+$ state in our results.
Our model is thus consistent with the long-life property of the $9/2_1^+$ state
in $^{69}$As as in $^{67}$As \cite{Hase3}.
   The states $13/2_2^+$ and $17/2_3^+$ have large neutron occupation numbers
$\langle n_{g9/2}^\nu \rangle = 0.61$
and $\langle n_{g9/2}^\nu \rangle = 0.68$, respectively.
They have structure similar to the $9/2_1^+$ state and are possibly members of
the band 1 on $9/2_1^+$.
We note here that calculated $Q_{\rm sp}$ values are negative (prolate)
for positive-parity states.

   The trend that a neutron is apt to jump in the $g_{9/2}$ orbit
manifests itself clearly in other states with 
$11/2^+ \le J^\pi \le 23/2^+$ and in the $25/2_3^+$ state.
These states have one neutron instead of one proton in the $g_{9/2}$ orbit
(see Fig. \ref{fig2}) and can be called ``high-$j$ $1n$" states.
We can say that the backbending from $21/2_1^+$ to $25/2_1^+$ in $^{69}$As
is caused by the ``$1p1n$ alignment" in contrast to the discussion in Ref.
\cite{Bruce}.
 This is unexpected structure in odd-proton nuclei.
The results suggest that the residual $A=68$ system excluding one $g_{9/2}$
nucleon favors an odd-odd subsystem with $T_{pf}=1$ analogous to $^{68}$As
rather than an even-even subsystem with $T_{pf}=2$ analogous to $^{68}$Ge. 
Because there is no energy difference between $1p$ and $1n$
in the $g_{9/2}$ orbit, the $A=68$ subsystems with $T_{pf}=2$ and $T_{pf}=1$
compete with each other. 
It should be noted here that the $T=0$ monopole field $H_{\pi \nu}^{T=0}$,
which brings about the bulk of the symmetry energy
depending on the total isospin \cite{Kaneko2},
does not have any effect on the subsystem but operates on the total isospin
of the whole system.
 The $A=68$ subsystem with
 $\langle J_{pf}\rangle \approx 0$ and $\langle T_{pf}\rangle \approx 2$
is probably superior in energy for the $9/2_1^+$ state.
For the $J^\pi >9/2^+$ states excluding $13/2_2^+$ and $17/2_3^+$, however,
the $A=68$ subsystem with $T_{pf} \approx 1$
can increase angular momentum with less energy.
This is the reason why the yrast states with $11/2^+ \le J^\pi \le 23/2^+$
have mainly one neutron in the $g_{9/2}$ orbit, in our shell model.
 The same mechanism probably produces such 
``high-$j$ $1n$" states in odd-proton $N \approx Z$ nuclei.

   Our calculations predict another configuration
for the yrast states $35/2_1^+$, $39/2_1^+$, and $43/2_1^+$.
These states have the expectation values
 $\langle n_{g9/2}^\pi \rangle = 1.7 \sim 1.5$,
 $\langle n_{g9/2}^\nu \rangle = 1.4 \sim 1.6$,
 $\langle J_{g9/2}\rangle \approx 25/2$,
 and $\langle T_{g9/2}\rangle \approx 1/2$,
 which suggests that these yrast states are mixed states of
the $2p1n$ and $1p2n$ aligned states. It is interesting that the $2p1n$
alignment contributes significantly to the unfavored states with $J^\pi=(7/2+2l)^+$.
 No detection of the $J^\pi=(7/2+2l)^+$ (unfavored) band at spins
$J^\pi \ge 31/2^+$ in experiment \cite{Stefanescu} seems to be related to
the mixed structure of the $J^\pi=(7/2+2l)^+$ band
 different from the $J^\pi=(9/2+2l)^+$ (favored) band.

\subsection{Negative-parity states}\label{sec3.2}

\begin{figure}[t]
\includegraphics[width=8.6cm,height=12.0cm]{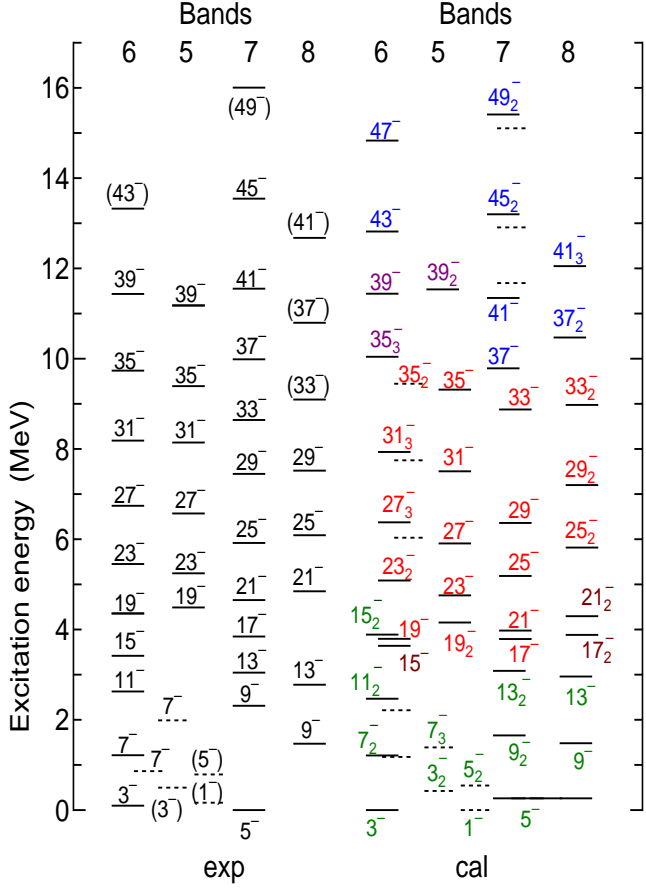}
  \caption{Experimental and calculated energy levels with negative parity,
           illustrated in the same manner as Fig. 1.}
  \label{fig3}
\end{figure}

   Let us turn our attention to negative-parity states. Calculated
negative-parity energy levels are compared with the experimental ones
\cite{Stefanescu} in Fig. \ref{fig3}, where the experimental bands
are labeled with the same numbers as those used in Ref. \cite{Stefanescu}
and corresponding theoretical bands are classified
according to cascade series of large $B(E2:J \rightarrow J - 2)$ values.
The calculated level scheme shows quantitative deviations from the
experimental one \cite{Stefanescu} in some details.
The deviations suggest adjusting finely the parameters of the model
Hamiltonian or introducing effects excluded from it.
 However, a lot of experimental energy levels are qualitatively well
reproduced except that the calculation lays the collective states
($3/2_1^-$ and $5/2_1^-$) and ($15/2_2^-$ and $19/2_1^-$) in reverse order. 
The calculated four bands are basically in agreement with the experimental ones.

\begin{figure}[t]
\includegraphics[width=8.2cm,height=11.5cm]{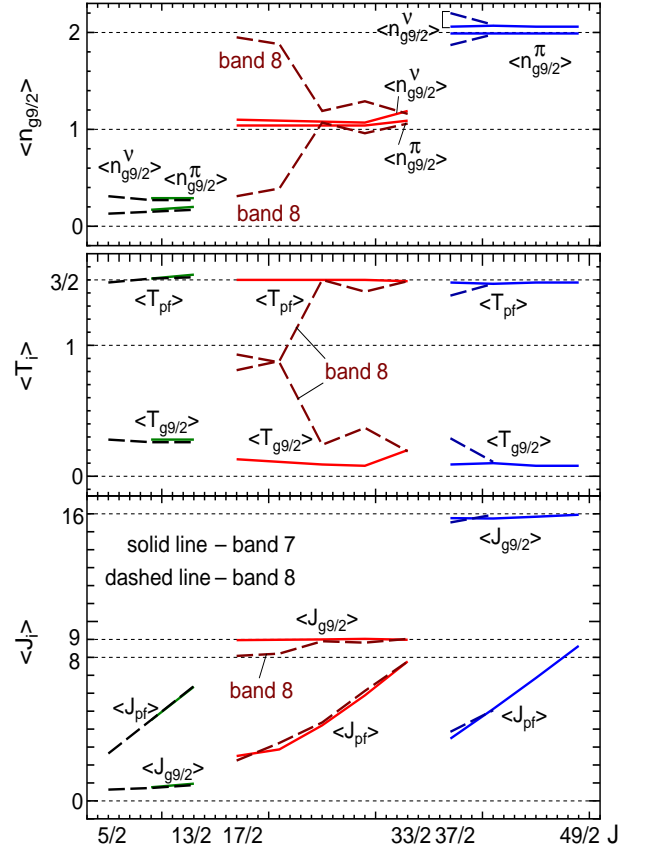}
  \caption{The expectation values $\langle n^\nu _{g9/2} \rangle$,
           $\langle n^\pi _{g9/2} \rangle$,
           $\langle T_i \rangle$, and $\langle J_i \rangle$
           ($i=g_{9/2},pf$) for the negative-parity bands 7 and 8.}
  \label{fig4}
\end{figure}

  Our shell model shows that, in a parallel manner to 
the positive-parity states, each of the negative-parity bands can be divided
into three stages A, B, and C:
The occupation number in the $g_{9/2}$ orbit is 
$\langle n_{g9/2}^\pi \rangle + \langle n_{g9/2}^\nu \rangle <0.57$
at the stage A,
$\langle n_{g9/2}^\pi \rangle + \langle n_{g9/2}^\nu \rangle \sim 2$
at the stage B, and 
$\langle n_{g9/2}^\pi \rangle + \langle n_{g9/2}^\nu \rangle \sim 4$
at the stage C.
(Note that a basis state with negative parity has an even number of
 $g_{9/2}$  nucleons.)
The three stages of each band are distinguished by illustrating them
at somewhat shifted columns in Fig. \ref{fig3}.
In Fig. \ref{fig4}, by way of example, we illustrate how the bands 7 and 8
change the expectation values $\langle n_{g9/2} \rangle$
and $\langle J_{g9/2} \rangle$ at the three stages.
It is reasonable to separate a negative-parity band into the three sub-bands
from the parallelism between Figs. \ref{fig1} (\ref{fig2})
and \ref{fig3} (\ref{fig4}).
Let us call the three sub-bands, for instance, 7A, 7B, and 7C.

The low-spin states with $J^\pi \le 7/2^-$ below $9/2_1^+$ have
only fractional numbers of nucleons in the $g_{9/2}$ orbit,
 $\langle n^\pi_{g9/2} \rangle <0.20$ and $\langle n^\nu_{g9/2} \rangle < 0.33$.
The occupation numbers in the three orbits ($p_{3/2}$,$f_{5/2}$,$p_{1/2}$)
show that these low-spin states are collective states strongly mixed in the
$pf$ space. 
This is in disagreement with the conjecture in Ref. \cite{Stefanescu}
 that the low-lying negative-parity states have single-particle character.
 Our model lays the collective state $3/2_1^-$
slightly below the $5/2_1^-$ state for the odd-proton $N \approx Z$ nuclei
 $^{69}$As and $^{67}$As \cite{Hase3},
which is inconsistent with the experimental ground state $5/2^-$.
However, our model reproduces the correct ground states $3/2^-$ and $1/2^-$
respectively for the odd-neutron $N \approx Z$ nuclei
$^{65}$Ge and $^{67}$Ge \cite{Hase2}.
In Fig. 3, tentatively assigned $1/2_1^-$, $3/2_2^-$, and
$5/2_2^-$ states \cite{ENSDF} are considerably well reproduced
with our model.
Repeated calculations by changing the parameters within the extended
$P+QQ$ Hamiltonian have not succeeded to get a ${5/2}^-$ ground state
in odd-mass As isotopes.
The reversed order of $5/2_1^-$ and $3/2_1^-$ suggests an effect
missing in our model.  There is a possibility that the imposed isospin
symmetry is too severe for reproducing different ground states
in the $N=33$ nucleus $^{65}$Ge and the $Z=33$ nuclei ($^{67}$As, $^{69}$As).

\begin{table}[b]
\caption{Calculated spectroscopic quadrupole moments $Q_{\rm sp}$ 
         in $e$ fm$^2$ for the lowest two sub-bands 6A and 8A.}
\begin{tabular}{c|cccc}   \hline
  6A    & $3/2_1^-$ & $7/2_2^-$ & $11/2_2^-$ & $15/2_2^-$  \\ \hline
 $Q_{\rm sp}$ & 22  &  18       &   41       &  13     \\ \hline \hline
  8A    & $5/2_1^-$ & $9/2_1^-$ & $13/2_1^-$ &             \\ \hline
 $Q_{\rm sp}$ & $-30$ &  $-19$  &  $-17$     &       \\ \hline
\end{tabular}
\label{table1}
\end{table}

\begin{figure}
\includegraphics[width=5.2cm,height=4.533cm]{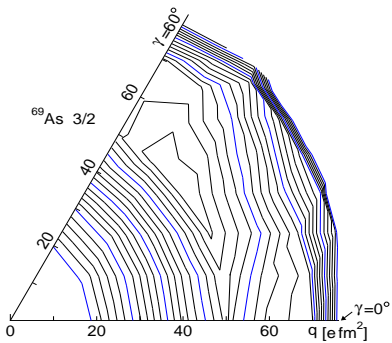}
  \caption{The energy surface
           $ \langle q, \gamma | H | q, \gamma \rangle $
           in the $q-\gamma$ plane ($0^\circ \le \gamma \le 60^\circ$)
           plotted with contours.}
  \label{fig5}
\end{figure}

   It is interesting that in calculated results the lowest sub-bands
6A ($3/2_1^-$,$7/2_2^-$,$11/2_2^-$,$15/2_2^-$)
 and 8A ($5/2_1^-$,$9/2_1^-$,$13/2_1^-$)
have distinctly positive and negative signs of spectroscopic quadrupole moments 
$Q_{\rm sp}$ as shown in Table \ref{table1}.  The shell model calculations
using the present model \cite{Hase1,Kaneko1,Hase2,Hase3} give $\pm$ signs
of $Q_{\rm sp}$ to the $2_1^+$ and $2_2^+$ states in even-even Zn and Ge
isotopes with $N \approx Z$ and to the $1_1^+$ and $3_1^+$ states in $^{66}$As,
but do not produce distinct two bands with $Q_{\rm sp}=\pm$ for these nuclei.
Only for $^{68}$Se, in our calculations, we got oblate-prolate shape coexistence
as two bands with $Q_{\rm sp}=\pm$. (That is confirmed by the two minima
of the potential energy surface drawn in the $q-\gamma$ plane
in the constrained HF calculation for the same Hamiltonian \cite{Kaneko1}.)
Therefore, the coexistence of the two bands with $Q_{\rm sp}=\pm$ in an odd-mass
$N \approx Z$ nucleus is remarkable. It is known, however, that
in this mass region the potential energy mostly has minima
both in prolate and oblate shapes under the condition of axial symmetry.
 In order to determine the shapes of the pair
of $Q_{\rm sp}=\pm$ bands, we must examine the potential energy surface
in the $q-\gamma$ plane. We carried out the constrained HF calculation
\cite{Mizusaki3,Hara} for $^{69}$As.
The result is illustrated in Fig. \ref{fig5},
where $q$ is the intrinsic quadrupole moment (the unit is $e \mbox{fm}^2$).
We have the relation $Q_{sp}\propto q(e_\pi + e_\nu)/e$ but cannot easily
express its coefficient in the triaxial situation shown in Fig. \ref{fig5}.
  This figure shows that in low energy
$^{69}$As favors a triaxial shape in the $q-\gamma$ plane but the shallow and
broad minimum indicates rather $\gamma$-unstable nature.
Thus, the coexistence of the $Q_{\rm sp}=\pm$ bands in $^{69}$As is different
from the oblate-prolate shape coexistence in $^{68}$Se but is a unique phenomenon
in an odd-mass nucleus.  The last odd nucleon must play an important role
in kinematically determining the nuclear shape.
  The two angular momentum
couplings $J=5/2+2l$ (favored) and $J=3/2+2l$ (unfavored)
seem to have different effects on $Q_{\rm sp}$.

   For high-spin negative-parity states, only $2n$, $1p3n$, and $2p2n$
configurations in the $g_{9/2}$ orbit are considered in Ref. \cite{Stefanescu}.
In contrast Fig. \ref{fig4} displays that unexpected structure appears
in the sub-bands 7B and 8B with $17/2^- \le J^\pi \le 33/2^-$.
The present shell model predicts the $1p1n$ configuration in the $g_{9/2}$ orbit
($\langle n_{g9/2}^\pi \rangle \approx 1$ and
 $\langle n_{g9/2}^\nu \rangle \approx 1$)
 for the states of the sub-bands 7B and 8B (also for the states of
 the sub-bands 5B and 6B with $19/2^- \le J^\pi \le 35/2^-$)
 except for the states $17/2_2^-$ and $21/2_2^-$.
The $1p1n$ pair in the $g_{9/2}$ orbit has
 $\langle T_{g9/2}\rangle \approx 0$ and $\langle J_{g9/2}\rangle \approx 9$
as shown in the lower panel of Fig. \ref{fig4},
 which indicates the $T=0$, $J=9$ $1p1n$ alignment in the high-$j$ orbit $g_{9/2}$.
The shell model calculations have already shown the same $1p1n$ alignment
in even-even and odd-odd $N \approx Z$ nuclei ($^{62}$Zn, $^{64-68}$Ge,
$^{68}$Se, and $^{66}$As) \cite{Hase1,Hase2,Hase3}.
Therefore we can say that the $T=0$, $J=9$ $1p1n$ alignment is a general
phenomenon in $N \approx Z$ nuclei of this mass region.
We can expect the same structure also in $^{72}$Kr, $^{72}$Br, $^{73}$Br,
 {\it etc}.

   When the $T=0$, $J=9$ $1p1n$ alignment takes place in $^{69}$As,
the residual $A=67$ subsystem excluding the $g_{9/2}$ nucleons has the isospin
$T_{pf}=3/2$ analogous to $^{67}$Ge (see Fig. \ref{fig4}).
The $T=0$ $1p1n$ alignment is not only due to a large energy gain of the 
$T=0$, $J=9$ $1p1n$ pair in the $g_{9/2}$ orbit but also due to 
low energy of the residual $A=67$ subsystem.
The competition between the $A=67$ subsystems with $T_{pf}=3/2$ and $T_{pf}=1/2$
affects the competition between the $T=0$ $1p1n$ alignment
and $T=1$ $2n$ alignment in the $g_{9/2}$ orbit.
 In our calculations, the $2n$ aligned state $17/2_2^-$ ($21/2_2^-$) is higher
in energy than the $1p1n$ aligned state $17/2_1^-$ ($21/2_1^-$),
 and the $2n$ aligned state does not appear in the lowest two states of each $J$
when $23/2^- \le J^\pi \le 35/2^-$.
This situation possibly explains the reason why the experiment \cite{Stefanescu}
detected only one $17/2^-$ state in the doublet of $J^\pi=(5/2+2l)^-$ bands.
The results $\langle n_{g9/2}^\pi \rangle \approx 0.67$ and
 $\langle n_{g9/2}^\nu \rangle \approx 1.57$ for $15/2_1^-$ suggest that
the $15/2_1^-$ state is a mixture of the $1p1n$ and $2n$ aligned states.

   In our calculations, the $J^\pi \ge 37/2^-$ states have
the aligned $2p2n$ in the $g_{9/2}$ orbit except that the states
$39/2_1^-$ and $39/2_2^-$ have mixed components of the aligned $1p3n$.
This is consistent with the result of cranked Nilsson-Strutinsky calculations
in Ref. \cite{Stefanescu}.
The calculated $35/2_3^-$ state is the $2p2n$ aligned state
mixed with the $1p3n$ aligned state and is a candidate for the $35/2^-$ member
of the band 6.

\section{Conclusion}\label{sec4}

   The analysis of $^{69}$As by means of large-scale shell model calculations
has revealed unexpected features of structure. In this odd-proton $N \approx Z$
nucleus in which the high-$j$ intruder orbit $g_{9/2}$ plays an important role,
one neutron competes with one proton for occupying the $g_{9/2}$ orbit
in the states $9/2_1^+$, $13/2_2^+$, and $17/2_3^+$. Moreover, one neutron
almost solely occupies the $g_{9/2}$ orbit in other positive-parity states
with $J^\pi \le 23/2^+$. This trend makes higher-spin states be the $1p2n$
and $2p3n$ aligned states.
For the negative-parity states, the present shell model indicates
the $T=0$, $J=9$ $1p1n$ alignment in the $g_{9/2}$ orbit,
in an odd-mass nucleus as well as odd-odd and even-even nuclei.
These unique configurations take place in cooperation with the characteristic
that the $N \approx Z$ subsystems with different $T$ excluding the $g_{9/2}$
nucleons nearly degenerate in energy.
The present model also predicts the coexistence of two lowest bands
with $Q_{sp}=\pm$ different from the known oblate-prolate shape coexistence,
in the odd-mass nucleus $^{69}$As.
This work and a series of our papers \cite{Hase1,Kaneko1,Hase2,Hase3}
clarified a variety of interesting phenomena in $N \approx Z$ nuclei,
providing a useful perspective for studying neighboring nuclei.
 



\end{document}